\documentclass[twocolumn,showpacs,preprintnumbers,amsmath,amssymb,superscriptaddress]{revtex4}
\usepackage{amssymb}
\usepackage{mathrsfs}
\usepackage{graphicx}
\usepackage{dcolumn}

\begin{document}
\title{Detecting Structure of Complex Network by Quantum Bosonic Dynamics \\}
\author{Xin Jiang}
\affiliation{Key Laboratory of Mathematics, Informatics and
Behavioral Semantics, Ministry of Education, Beijing University of
Aeronautics and Astronautics, 100191 Beijing, China}

\author{Hailong Wang}
\affiliation{Key Laboratory of Mathematics, Informatics and
Behavioral Semantics, Ministry of Education, Beijing University of
Aeronautics and Astronautics, 100191 Beijing, China}
\affiliation{School of Physics,Peking University, 100871 Beijing,
China}

\author{Lili Ma}
\affiliation{Key Laboratory of Mathematics, Informatics and
Behavioral Semantics, Ministry of Education, Beijing University of
Aeronautics and Astronautics, 100191 Beijing, China}

\author{Zhanli Zhang}
\affiliation{School of Mathematical Sciences, Peking University,
100871 Beijing, China}

\author{Shaoting Tang}
\affiliation{Key Laboratory of Mathematics, Informatics and
Behavioral Semantics, Ministry of Education, Beijing University of
Aeronautics and Astronautics, 100191 Beijing, China}

\author{Guangshan Tian}
\affiliation{School of Physics,Peking University, 100871 Beijing,
China}

\author{Zhiming Zheng}
\email{zzheng@pku.edu.cn}
\affiliation{Key Laboratory of Mathematics, Informatics and
Behavioral Semantics, Ministry of Education, Beijing University of
Aeronautics and Astronautics, 100191 Beijing, China}

\date{\today}

\begin{abstract}

  We introduce a non-interacting boson model to investigate
topological structure of complex networks in the present
paper. By exactly solving this model, we show that
it provides a powerful analytical tool
in uncovering the important properties of real-world
networks. We find that the ground state degeneracy
of this model is equal to the number of connected components
in the network and the square of coefficients in the expansion
of ground state gives the averaged time for a random
walker spending at each node in the infinite
time limit. Furthermore, the first excited state
appears always on its largest connected component.
To show usefulness of this approach in practice,
we carry on also numerical simulations on some concrete
complex networks. Our results are completely consistent
with the previous conclusions derived by graph theory
methods.

\end{abstract}

% \keywords{Suggested keywords}
% Use showkeys class option if keyword
% display desired
\maketitle

  Currently, the study of complex networks has become
an important research field in physics, biology,
sociology, information technology and other branch
of sciences \cite{Albert,Newman,Dorogovtsev}.
A characteristic feature of these systems is the stochastic diffusion
of some discrete objects on them. Intuitively, such process is
referred to as granular flows. Perhaps the most well-known example
is the flow of information on Internet. In this case, messages are
encapsulated into discrete date packets to be sent from one computer
to another. It has been observed that, despite randomness of each
single packet hopping, the motion of massive packets, governed by
network protocols may reach certain nonequilibrium steady states,
such as self-similarity of the Ethernet traffic \cite{Leland}.
An important issue arose is whether some general principles can be
found to describe dynamics of granular flows in a given complex
network \cite{Tenleadings}. Various techniques have been developed
to address this problem
\cite{Noh1,Danila,Wang,Stauffer,Baronchelli,Noh2,Maragakis,Evans,Lopez,Crovella,Moura,Germano}.
For instance, in Refs.~\cite{Moura,Germano}, both itinerant fermion
and boson models were introduced to study the distribution of
information packets on Internet.

  On the other hand, since granular flow
on a complex network is completely determined by its topology
and protocols, one would naturally like to ask whether knowledge
on dynamics of these particles can be conversely used to detect
structure of the network, such as its connectedness as well as
the number of its components. Previously, this problem has been studied
in framework of percolation theory \cite{Callaway}, epidemiological
processes \cite{Satorras}, and network search \cite{Watts}.
For example, the authors of Ref.~\cite{Goltsev} has shown
that spectrum of the branching matrix can be used
to uncover properties of the largest connected component
of a network. However, a unified approach is still lacking.

  In this paper, we shall introduce an itinerant bosonic model,
whose hopping amplitude between any pair of nodes is determined
by the local topology of networks. As we show,
information on their global properties can be derived
from dynamics of this quantum system.
In particular, we find that the ground state of this model
gives the static distribution of particles
and its degeneracy accounts for the number of connected
components in the networks. Furthermore, we are able to identify
the largest connected component by calculating
the energy gap between the first excited state and the ground state
of the system.

  To begin with, let us consider a large but finite complex network
with $N$ nodes and a set of links connecting them.
Define the the network adjacency matrix $A$ by the following rules:
If nodes $i$ and $j$ are connected by a link, the matrix element
$A_{ij}$ is set to be unit; Otherwise, $A_{ij}=0$. A particle
(information packet) can be transmitted between two nodes
if they are linked. Furthermore, we assume that
the network is undirected. In this case, $A^T=A$ holds true.
With these definitions, the degree for a specific
node $i$ is given by $K_i=\sum_{l\ne i}A_{il}$.

  For such a network, one would first like to know
whether it is connected and which component is the largest
one, if it is disconnected \cite{Bollobas}.
To answer these questions, we consider a non-interacting
boson model, whose Hamiltonian is given by
\begin{equation}
\hat{H} = \sum_{\{i,j\}} t_{ij} \hat{c}_i^\dagger\hat{c}_j,
\label{Hamiltonian}
\end{equation}
where $\hat{c}_i^\dagger$ ($\hat{c}_i$) denotes the
boson creation (annihilation) operator which creates (annihilates)
a spinless boson at node $i$. $t_{ij}$ represents
the boson hopping amplitude between nodes $i$ and $j$
and the summation is over all the distinct pairs of nodes.

  To build local topology of the network into this model,
we further define $t_{ij}$ by
\begin{eqnarray}
t_{ij} = \left\{
\begin{array}{ll}
- A_{ij}, & {\rm if}\>i\ne j; \\
\frac{\sum_{l\ne i} A_{il} \sqrt{K_l}}{\sqrt{K_i}},
& {\rm if}\>i = j.
\end{array}
\right.
\label{Hopping Amplitude}
\end{eqnarray}
While the first line in Eq.~(\ref{Hopping Amplitude})
seems natural, the second one demands some explanation.
In physics, $t_{ii}$ is the local chemical potential
of particles. It controls the probability for a particle
to stay at node $i$: A higher value of $t_{ii}$ makes
such probability lower. On the other hand, for a particle
hopping randomly in network, large value of $\sum_{l\ne i}A_{il}\sqrt{K_l}$
implies that it has less chance to return to node
$i$. Therefore, visibility of the particle at this node
is decreased. It justifies our choice for the numerator
of $t_{ii}$. In the meantime, we introduce denominator
$\sqrt{K_i}$ in $t_{ii}$ for the purpose of normalization.

  We notice that a similar model has been recently introduced
to study localization of light wave-packet in complex
networks \cite{Jahnke}.

  Now, by choosing a natural basis of single-particle
states $\vert i\rangle=\hat{c}_i^\dagger\vert 0\rangle$,
we re-write Hamiltonian (\ref{Hamiltonian})
into the following matrix
\begin{equation}
H = \left(
\begin{array}{cccc}
\frac{{\sum}^\prime_l A_{1l}\sqrt{K_l}}{\sqrt{K_{1}}} &
- A_{12} &\ldots &\ldots \\
- A_{21} & \frac{{\sum}^\prime_l A_{2l}\sqrt{K_l}}{\sqrt{K_{2}}}
& \ldots &\ldots \\
\ldots & \ldots & \ldots \\
- A_{N1} & - A_{N2} & \ldots
& \frac{{\sum}^\prime_l A_{Nl}\sqrt{K_l}}{\sqrt{K_{N}}}
\end{array}
\right),
\label{Hamiltonian Matrix}
\end{equation}
where ${\sum}^\prime_l$ stands for $\sum_{l\ne i}$.
We will show that the ground state and the first excited
state of this model provide us with information on the network
structure.

  First, we notice that matrix $H$ is semi-positive definite.
In fact, by making a similar transformation
with $K=diag(\sqrt{K_{1}},\sqrt{K_{2}},...\sqrt{K_{N}})$,
we obtain
\begin{eqnarray}
& &
\widetilde{H} = K^\dagger HK \nonumber \\
& = &
\left(
\begin{array}{ccc}
{\sum}^\prime_l \sqrt{K_1K_l}A_{1l} & \ldots & - \sqrt{K_1K_N} A_{1N} \\
- \sqrt{K_2K_1} A_{21} & \ldots & - \sqrt{K_2K_N} A_{2N}  \\
\ldots & \ldots & \ldots \\
- \sqrt{K_NK_1} A_{N1} & \ldots & {\sum}^\prime_l \sqrt{K_NK_l} A_{Nl}
\end{array}
\right).
\end{eqnarray}
To this matrix, we can apply Gershgorin's theorem \cite{Franklin},
which tells us that, for a $N\times N$
hermitian matrix $A$, each of its eigenvalues must satisfy,
at least, one of the following inequalities
\begin{equation}
\vert\lambda - a_{mm}\vert
\le \sum_{n\ne m}\vert a_{mn}\vert,\>\>\>m=1,2,\cdots,N.
\label{Gershgorin Inequality}
\end{equation}
Since $a_{mm}={\sum}^\prime_n\vert a_{mn}\vert$ holds
for each row index $m$ in $\widetilde{H}$, we conclude immediately
by applying Eq.~(\ref{Gershgorin Inequality}) that any eigenvalue
of the matrix is bounded below by zero. Therefore,
$\widetilde{H}$ as well as $H$ are semi-positive definite.

  On the other hand, a direct calculation reveals that vector
${\bf u}=(\sqrt{K_1},\sqrt{K_2},\cdots,\sqrt{K_N})^T$
is an eigenvector of matrix $H$ with eigenvalue $\lambda=0$.
Therefore, we find that
\begin{equation}
\Phi_0 = \frac{1}{\sqrt{\sum_n K_n}}\sum_i\sqrt{K_i}\vert i\rangle
\end{equation}
is one of the normalized ground state of Hamiltonian (\ref{Hamiltonian}).
According to quantum mechanics, square of coefficient
$K_i/\sqrt{\sum_nK_n}$ equals the probability of finding a particle
at node $i$ in the ground state. This conclusion coincides with
a well-known result in graph theory that, in the static
state, the time of a randomly-walking particle spending
at node $i$ is proportional to $K_i$ \cite{Bollobas}.
It indicates that the ground state $\Phi_0$
of Hamiltonian (\ref{Hamiltonian}) represents actually
the static state of the corresponding complex network.

\begin{figure}
\includegraphics[width=0.51\textwidth]{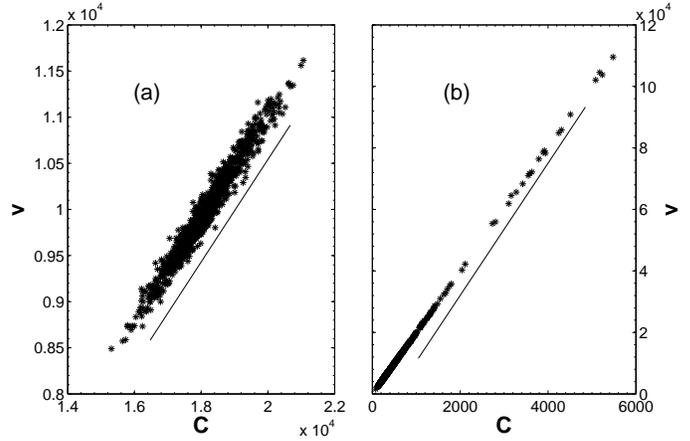}
\caption{\label{fig:center} Number of visitations $V$ vs $C$ in (a)
Erd\H{o}s-R\'{e}nyi networks with $N=1000$ nodes and the connecting
probability $p=0.36$ and (b) scale-free networks generated by the
Barab\'{a}si-Albert algorithm, with $N=1000$ nodes, and an average
degree of $6$. We assign a random value $w_{ij}$, which is randomly
chosen from [1,100], to each $t_{ij}$. We originally set $M=5$ and
$N_{s}=10^{10}$ to ensure the system reach the equilibrium phase.
The simulations have at least $10^3$ realizations.}
\end{figure}

  Also, this approach can be easily applied
to more general cases, in which nodes are linked
with unequal weight $\{w_{ij}\}$. To deal with those cases,
we multiply each link $A_{ij}$ by $w_{ij}$
and then replace degree $K_i$ by  $C_i=\sum_{l\ne i}w_{il}A_{il}$,
a quantity which is similar to fitness defined in Ref.~\cite{Bianconi},
in Eq.~(\ref{Hopping Amplitude}). With these changes,
we can show that one of the ground state
of the Hamiltonian is given by
\begin{equation}
\widetilde{\Phi}_0 = \frac{1}{\sqrt{\sum_nC_n}}
\sum_i^N\sqrt{C_i}\vert i\rangle
\end{equation}
by repeating the above procedure. Therefore,
the probability for finding a particle
at node $i$ should be equal to
\begin{equation}
p_i = \frac{C_i}{\sum_n C_n}
\end{equation}
in the infinite time limit.

  Our numerical simulation confirms
also these results. By displacing randomly $M$ non-interacting
particles on random scale-free networks with a specific
set of linking amplitudes, we count the number $v_i$ of visitations
to node $i$ after each particle hops $N$ steps.
As shown in Fig.~1(b), indeed, $v_i$ is strictly proportional
to capacity $C_i$ of the node. Furthermore, we find that similar
results for ER networks[Fig.1(a)].

  Up to now, we have not discussed degeneracy of the ground state yet.
In fact, this quantity gives the number of connected components
in the network. To make this point more clear, let us first consider
a completely connected network. In this case, matrix $H$
in Eq.~(\ref{Hamiltonian Matrix}) is irreducible.
Furthermore, its off-diagonal elements are either zero
or negative unit. Therefore, Perron-Fr\"obenius theorem
in matrix theory applies \cite{Franklin}. It tells us that
the lowest eigenvector of the matrix is unique. In other words,
the ground state $\Phi_0$ is nondegenerate in this case.

  On the other hand, if the network consists of several disjoint
components, then Hamiltonian (\ref{Hamiltonian})
has a nondegenerate ground state on each of them.
Moreover, as shown above, the lowest
eigenvalue of the Hamiltonian is always zero. Consequently,
degeneracy of the ground-state energy $E_0=0$ is equal
to $N_c$, the number of connected components in the network.

  To show usefulness of this result, we investigate numerically
some widely studied complex network, the so-called
Poisson random graph. In this model, the probability
for degree $K_i$ of node $i$ taking on a specific value $k$
is given by
\begin{eqnarray}
p_k = e^{-c}\frac{c^k}{k!},
\end{eqnarray}
where $c>0$ is the distribution mean. In particular,
we consider an ensemble of graphs. Each of them
has $n$ nodes and $m$ edges and appears with equal probability.
In the following, we denote this ensemble by $G_{n,m}$.

  Our data are shown in Fig.~2. We find that, for $m\le n/2$,
degeneracy of the ground-state is roughly equal to $D=n-m$.
However, when $z=2m/n>1$, $D$ decreases non-linearly as $m$
further increases. This change of behavior around $z=1$
can be clearly seen by taking the first order derivative
of the fitting curves. It suggests that a dynamic phase transition
may occur at point $z=1$. Our findings are consistent
with the previous graph theoretic results \cite{Bollobas}.

\begin{figure}
\includegraphics[width=0.49\textwidth]{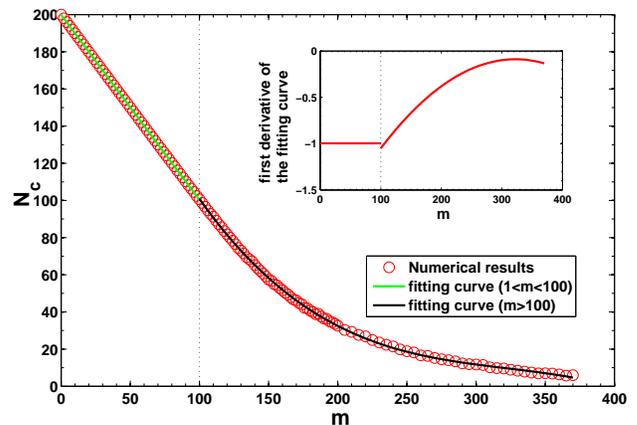}
\caption{\label{fig:center}The number of components in random graphs
with $n$ vertices and $m$ edges. Each point is an average over
$10^3$ networks of $200$ vertices each.}
\end{figure}

  In addition, we observe also that the first excited state
of Hamiltonian (\ref{Hamiltonian}) is generally nondegenerate
in our simulations. Moreover, its wave function
is actually located on the the largest connected component
of the network. In other words, the coefficients $b_i$'s
in the expansion of the first excited wave function
\begin{equation}
\Phi_1 = \sum_i b_i \vert i\rangle
\end{equation}
are only nonzero for those nodes which belong to the same connected
subset of the largest size. Our data are shown in Fig.~3. From it,
one can easily see that, as $m$ varies, the number of nonzero
expansion coefficients in $\Phi_1$ is always equal to the size of
the largest connected component, which is figured out by the
standard depth-first search traversing algorithm. Moreover, there
always exists a single largest connected component in each
realization of network.

  From quantum mechanical point of view, this observation
is not difficult to understand. Let us think of each connected
component of the network as a three-dimensional infinitely deep
potential well. Then, as is well known in quantum mechanics,
the energy gap between the first excited state
and the ground state of Hamiltonian (\ref{Hamiltonian})
in each well is proportional to $V^{-2/3}$, where $V$
is volume of the well \cite{Landau}.
In other words, the gap decreases as the number of nodes
in the connected component becomes larger.
Consequently, one will expect that the smallest energy gap
is reached in the largest well (connected component). On the other hand,
we have shown above that all the ground state energies
on the connected components of network are fixed at
$E_0=0$. Therefore, the global first excited state
must appear in the largest connected component.

\begin{figure}
\includegraphics[width=0.45\textwidth]{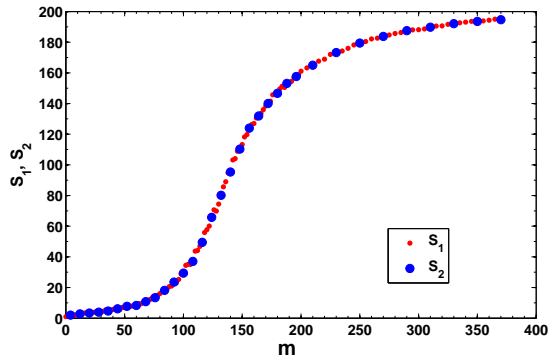}
\caption{\label{fig:center}The largest component size $S_{1}$ and
the number of $\Phi_1$'s nonzero expansion coefficients $S_{2}$ in
ER random networks with $n=200$ vertices and $m$ edges. Red points
indicate the number of nonzero expansion coefficients in $\Phi_1$,
while blue points are the results of the depth-first search
traversing algorithm on the same network. Each point is an average
over $10^3$ networks of $200$ vertices each.}
\end{figure}

  Furthermore, we conjecture that the higher excited
energy spectrum of this model can be used to reveal
other characters of networks, such as modularity.
We shall address this issue in future investigation.

  In summary, we introduce a non-interacting boson model
on complex networks to investigate their structures
in the present paper. Our motivation is based on the following
observation: It is the structure of a specific network
that determines the flow of particles in it. Therefore, our knowledge
on particle dynamics should be conversely useful to uncover
topology of the network. Indeed, we show that the ground
state energy of this model is degenerate if the network
consists of $N_c$ disjoint connected components
and its degeneracy coincides with $N_c$. Moreover,
the square of each coefficient in the expansion of ground
state gives the correct averaged time for a random
walker spending at a specific node $i$ in the infinite
time limit. Finally, we show also that the first
excited state of the model is always supported
on the largest connected component of network.
Therefore, it gives us a practical way to detect topological
structures of complex networks.

  This work is partially supported by the National Basic Research
Program of China (Grant No. 2005CB321900). One of us (G. S. T) is
also supported by the Chinese National Science Foundation under
Grant No. 10674003 and MOST grant 2006CB921300.

\end{document}